\documentclass[12pt]{iopart}

\usepackage{graphics}
\usepackage{amsthm}
\usepackage{graphicx}
\usepackage{verbatim}
\usepackage{subfigure}
\usepackage{hyperref}
\usepackage{epstopdf}
\usepackage{color}
\usepackage{epsfig}
\usepackage{amssymb}

\newcommand{\bea}{\begin{eqnarray}}
\newcommand{\eea}{\end{eqnarray}}
\newcommand{\ket}[1]{\left|{#1}\right\rangle}
\newcommand{\bra}[1]{\left\langle{#1}\right|}

\newcommand{\modu}[1]{\left|{#1}\right|}

\newcommand{\trev}{T_{\rm rev}}

\begin{document}

\title{Entanglement dynamics of quantum states  in a  beam splitter}
\author{M. Rohith}
\address{Department of Physics, Indian Institute of Space Science and Technology,\\
 Thiruvananthapuram 695 547, India\\
 rohith.11@iist.ac.in}
\author{R. Rajeev}
\address{Department of Physics, Indian Institute of Technology Madras,\\
 Chennai 600 036, India\\
 rajeevriitm@gmail.com}
\author{C. Sudheesh}
\address{Department of Physics, Indian Institute of Space Science and Technology,\\
 Thiruvananthapuram 695 547, India\\
 sudheesh@iist.ac.in}

\maketitle

\begin{abstract}
We theoretically study the dynamics of entangled states created  in a beam splitter with  a  nonlinear Kerr
 medium placed into  one input arm.  Entanglement  dynamics of   initial  classical and  nonclassical  states are studied
  and compared. Signatures of revival and fractional revival phenomena   exhibited  during the time evolution of states in 
  the Kerr medium are captured in the entangled states produced by the beam splitter. Maximum entanglement is obtained  at 
  the instants of   collapses  of wave packets  in the medium.    Our analysis shows   increase in entanglement with increase
   in the degree of nonclassicality of the initial states considered. We  show  that the states  generated at the output of the beam splitter using initial nonclassical  states are  more robust against decoherence, due to photon absorption by an environment,  than those formed by an initial classical state.
   \end{abstract}

\section{Introduction}
Quantum entanglement plays a crucial role in quantum information and quantum computing. It has been a key resource for quantum 
information processing. After the celebrated EPR paper \cite{epr} a tremendous amount of work has been done in the field of
 quantum entanglement. A detailed review of quantum entanglement is given in \cite{ryszard}. In most of the quantum information 
 processes, such as, quantum teleportation \cite{bennett}, quantum cryptography \cite{gisin}, superdense coding \cite{bennet2}, 
 quantum metrology \cite{Giovannetti}, etc., the systems are prepared initially in an entangled state. 
  Much attention is devoted to the discussion of entanglement properties of continuous variable systems, for their great practical 
  relevance in applications to quantum optics and quantum information \cite{adesso}. A family of entanglement witnesses based on 
  continuous
variable local orthogonal observables  to detect and estimate entanglement of Gaussian and
non-Gaussian states are presented in \cite{zhang}.

A beam splitter generates entangled states if the input  fields are nonclassical \cite{kim}. It has been shown that a standard nonlinear optics interaction,
arising from  a Kerr nonlinearity, followed by a simple interaction with a beam splitter produces large amount of entanglement in an arbitrarily short time \cite{vanEnk}. Here, the initial state considered  is a coherent state  and the input state for the beam splitter are taken at specific instants (fractional revival times)  during the time  evolution of the coherent state in the Kerr medium.
Recent experimental observation of quantum state collapse and revival of an initial coherent state due to 
the single-photon Kerr effect  \cite{gerhard} opens  up new directions for continuous variable 
quantum computation.  Because the production of large Kerr nonlinearity are possible \cite{gerhard}, we   study, to a greater extent,   the continuous dynamics of entanglement  using the state at  any instants instead of  at specific instants,  during the  evolution of coherent state in  the Kerr medium using the set-up in \cite{vanEnk}. It would also be interesting to compare the entanglement dynamics of an initial nonclassical state with the initial coherent state (classical state). For this purpose we consider 
  $m$-photon-added coherent states \cite{agarwal} because 
single-photon-added coherent states are experimentally generated using parametric down conversion in a nonlinear crystal  and characterised by quantum tomography \cite{zavatta}. 
Entanglement generation  using beam splitter by injecting photon-added coherent state in one input mode and vacuum state in other one is theoretically studied  in  \cite{berrada}. 
We also consider the impact of noise upon the entangled states generated in the beam splitter. This has been done by analysing  the effect of decoherence   due to photon absorption by an environment on the entangled states.

\begin{figure}[htpb]
\centering
\includegraphics[scale=0.5]{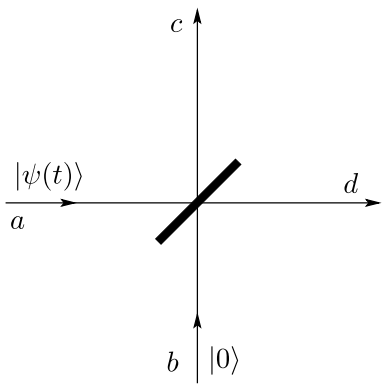}
\caption{A $50/50$ beam splitter with $\ket{\psi(t)}$ in the horizontal port and $\ket{0}$ in the vertical  port. $a$ and $b$ ($c$ and $d$) are the input (output) field modes respectively.}
\label{beamsplitter}
\end{figure}

The effective Hamiltonian for the propagation of single mode field in a Kerr  medium is given by \cite{milburn,kita}
	\begin{equation}
		H=\hbar \chi {a^\dag}^{2}a^{2}=\hbar \chi N(N-1) \label{Hamiltonian},
	\end{equation}
where $\chi$ is the third order nonlinear susceptibility of the medium, and $a^\dag$, $a$ are the photon creation and annihilation operators respectively. $N=a^\dag a$ is the number operator whose eigenstates are the Fock states, represented by $\ket{n}$. 
Consider the dynamics of an initial wave packet $\ket{\psi(0)}$ governed by the nonlinear Hamiltonian given in Eq. (\ref{Hamiltonian}). Following the Schr\"{o}dinger evolution 
\begin{equation}
\ket{\psi(t)}=\exp[-iHt/\hbar]\ket{\psi(0)},
\end{equation}
the state at any instant $t$ can be written as,
	\begin{equation}
		\ket{\psi(t)}=\exp[-i\chi tN(N-1)]\ket{\psi(0)}. \label{psit}
	\end{equation}
Here the interesting thing is that even if the initial wave packet is a classical one $\ket{\psi(t)}$  becomes nonclassical during the evolution in the Kerr medium \cite{yurke,tara,sudheesh1}. 
Subsequently we split the resulting nonclassical state    in a  beam splitter with the vacuum and generate entangled states.  
We consider a $50/50$ lossless beam splitter with $\pi/2$ phase difference between reflected and transmitted beam. A schematic representation of the  beam splitter is given in Fig. \ref{beamsplitter}. 
Here $a$ and $b$ ($c$ and $d$) are the input (output) field modes of the beam splitter respectively.  The state at a particular instant of time $t$, $\ket{\psi(t)}$, is achieved by adjusting the interaction length (time) of the medium. We inject the  field $\ket{\psi(t)}$  through the horizontal port (mode $a$) of the beam splitter and a vacuum state through the vertical port (mode $b$).   Thus the  input state  to the beam splitter is   $\ket{\psi(t)}_a\otimes\ket{0}_b$.  We omit the tensor product  symbol between the states in the rest of the paper. The state $\ket{\Phi(t)}$ of the output modes can be obtained using the unitary operator  $U$ of the beam splitter:
\begin{equation}
\ket{\Phi(t)}=U\,\big(\ket{\psi(t)}_a\ket{0}_b\big),
\end{equation}
where
	\begin{equation}   	
		U=\exp\left[i\frac{\pi}{4}(a^{\dag}b+ab^{\dag})\right]. \label{BS}
	\end{equation}
The state $\ket{\Phi(t)}$ is a two mode pure state of the field. The total density matrix for the state $\ket{\Phi(t)}$ is $\rho_{cd}=\ket{\Phi(t)}\bra{\Phi(t)}$. Amount of entanglement $E$, in such a pure bipartite state $\rho_{cd}$ can be calculated using the von Neumann entropy, which is defined as 
	\begin{equation}
		E=-Tr \left[\rho_r \log_2 \rho_r\right], 
		\label{entropy}
	\end{equation}
where $\rho_r$ ($r=$ mode $c$ or $d$) is the reduced density matrix of either of the subsystems $c$ or $d$.  The state $\ket{\Phi(t)}$ will be entangled whenever $\ket{\psi(t)}$ is a nonclassical state. 

The rest of the  paper is organised as follows. In Section 2,  we study the entanglement dynamics of an initial coherent state for various values of field strength.   Signatures of fractional revivals are manifested in   the von Neumann entropy and  arbitrary large amount of entanglement are generated at the instants of collapses of the initial coherent state.  In Section 3,  entanglement dynamics of initial $m-$photon-added coherent states are discussed and compared it with an initial coherent state. We find that    entanglement increases  with increase in the degree of nonclassicality of the photon-added coherent states.   
In Section 4, we checked the robustness of the entanglement states against the decoherence due to  photon absorption by an environment. Section 5 concludes  our main results.
 
\section{Entanglement dynamics of an initial coherent state} 
Consider the evolution of an initial coherent state $\ket{\alpha}$ through the  Kerr medium.   Let $\alpha=\nu^{1/2}\,\exp(i\theta)$, where $\nu=|\alpha|^2$ is the mean number of photons in the coherent state.  Without loss of generality, we set $\theta=\pi/4$ throughout this paper.
Substituting  $\ket{\psi(0)}=\ket{\alpha}$ in the Eq. (\ref{psit}), we get  
	\begin{equation}
		\ket{\psi(t)}=e^{-\modu{\alpha}^2/2}\sum_{n=0}^{\infty}\frac{\alpha^n e^{-i\chi t n(n-1)}}{\sqrt{n!}}\ket{n},\label{psitCS}
	\end{equation}
	where we have used the Fock state representation of the coherent state.
Collapses and revivals of wave packets are observed during the evolution of wave packets in the medium.   It can be shown that the state $\ket{\psi(t)}$ given in Eq. (\ref{psitCS}) revives periodically with revival time $\trev=\pi/\chi$. It also shows fractional revivals when the wave packet is split into a finite number of scaled copies of initial wave packet \cite{tara,miranowicz}.  

In between time  $t=0$ and $t=\trev$, $\ket{\psi(t)}$  shows $q$-sub-packet fractional revivals at times 
\begin{equation}
t=p\pi/q\chi, 
\label{frtime}
\end{equation}
where $p=1,2,...(q-1)$, for a given value of $q$ ($>1$) with the condition that $p$ and $q$ are mutually prime integers. (This is a necessary condition for the occurrence  of fractional revivals and other conditions for observation of revivals are discussed later.) The time evolved state at $t=\pi/q\chi$ is a superposition of $q$ coherent states:

\begin{equation}
\ket{\psi(\pi/q \chi)}=\left\{\begin{array}{ll}
\sum_{j=0}^{q-1} f_{j}\ket{\alpha \,e^{-2\pi ij/q}},\quad&
q \quad{\rm odd;} \\[10pt]
\sum_{j=0}^{q-1} g_{j}\ket{\alpha \,e^{i\pi/q}\,e^{-2\pi ij/q}},\quad&
q \quad{\rm even.} \end{array} \right.
\label{superpositionofcs}
\end{equation}
 where the coefficients $f_j$ and $g_j$ are known \cite{tara}. 
 If one subsequently takes these states and splits them on a 50/50 beam splitter with the 
 vacuum, the output state is an entangled state 
 
 \begin{equation}
\ket{\Phi_q}=\left\{\begin{array}{ll}
\sum_{j=0}^{q-1} f_{j}\ket{\beta \,e^{-2\pi ij/q}}_c\ket{i\beta \,e^{-2\pi ij/q}}_d,\quad&\\
\qquad\qquad\qquad \qquad\qquad\qquad \qquad (q \quad{\rm odd;})&\\[10pt]
\sum_{j=0}^{q-1} g_{j}\ket{\beta \,e^{i\pi/q}\,e^{-2\pi ij/q}}_c\ket{i\beta \,e^{i\pi/q}\,e^{-2\pi ij/q}}_d,\quad&\\
 \qquad\qquad\qquad \qquad\qquad\qquad\qquad (q \quad{\rm even;})&
 \end{array} \right.
\label{entangledcs}
\end{equation}
where $\beta=\alpha/\sqrt{2}$ \cite{vanEnk}. For large values of $\nu=|\alpha|^2$, the state 
$\ket{\Phi_q}$ is a maximally entangled state.

\begin{figure}[htpb]
\centering
\includegraphics[scale=0.47]{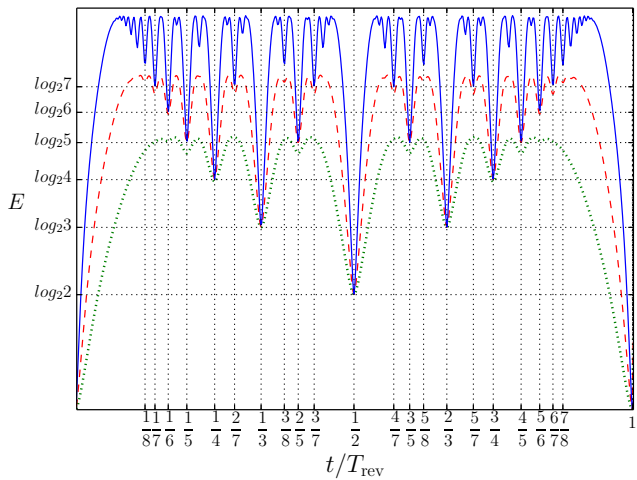}
\caption{Entanglement as a function of time $t/\trev$ from time $t=0$ to $t=\trev$ for an initial coherent state $\ket{\alpha}$ with $\modu{\alpha}^2=5$ (dotted), $10$ (dashed) and $20$ (solid). Local minima of the curves indicates the instants of fractional revivals.}
\label{m0nu510}
\end{figure}

Now we study the  entanglement at any time $t$ for the initial coherent state. Beam splitting action of the state $\ket{\psi(t)}$  given in  Eq. (\ref{psitCS}) with vacuum $\ket{0}$ results in the  state
	\begin{eqnarray}
		\ket{\Phi(t)}&=&e^{-\modu{\alpha}^2/2}
\sum_{n=0}^\infty \frac{\alpha^n}{\sqrt{n!}}\frac{\exp\left[-i\chi t n(n-1)\right]}{\quad 2^{n/2}} \nonumber\\
&\times&\sum_{p=0}^{n}{{n}\choose{p}}^{1/2}\ket{p}_c\ket{n-p}_d.\label{PhiCS}
	\end{eqnarray}
We numerically calculated the entanglement $E$ of the state $\ket{\Phi(t)}$ using Eq. (\ref{entropy})  and plot it  in Fig. \ref{m0nu510} for various values of $\nu=|\alpha|^2$ between the time  $t=0$ and $T_{\rm rev}$. 
It can be seen that at time $t=0$ the entanglement is zero because at this instant both the input states feeding into the beam splitter are classical states. The input state to the beam splitter is   $\ket{\alpha}_a \ket{0}_b$ and  the ouput state given in Eq. (\ref{PhiCS}) takes the form $\ket{\Phi(0)}=\ket{\frac{\alpha}{\sqrt{2}}}_c\ket{\frac{i\alpha}{\sqrt{2}}}_d$. This output state is  separable  and the  entanglement between the two output modes  $c$ and $d$ is  zero.   

During the time evolution in the medium, the state  $\ket{\psi(t)}$  exhibits nonclassical behaviour and the output state $\ket{\Phi(t)}$ shows non-zero entanglement. When the field
 strength $\nu=|\alpha|^2$ increases the entanglement increases in general and at the instants of fractional revivals it   takes  a local minima in the Fig \ref{m0nu510}.  Fractional revival times are marked in the figure with  vertical dotted lines. It should be noted that  R\'{e}nyi entropy takes local minima at fractional revival times in the single mode case where there is no  question of entanglement \cite{rohith}. The entanglement using von Neumann entropy  plot presented here shows   clear signatures of fractional revivals.   At $q$-sub-packet fractional revival times the output states given in Eq. (\ref{PhiCS}) reduces to the states given  Eq. (\ref{entangledcs}) which  are  maximally entangled states in $q$-dimension with $E=\log_2\,\, q$ for sufficiently large values of $\nu$ \cite{vanEnk}. It is evident from the figure that for $|\alpha|^2=5$ the state at $2, 3$ and $5$-sub-packet fractional revivals are maximally entangled state  with $E = \log_2 2, \log_2 3$ and $\log_2 4$, respectively. It can be verified from the figure  that the state at higher order fractional revival times  are maximally entangled states for larger values of  $\nu$. At revival times the  entanglement  returns to its initial value of zero.

\begin{figure*}[htpb]
\centering
\includegraphics[scale=0.6]{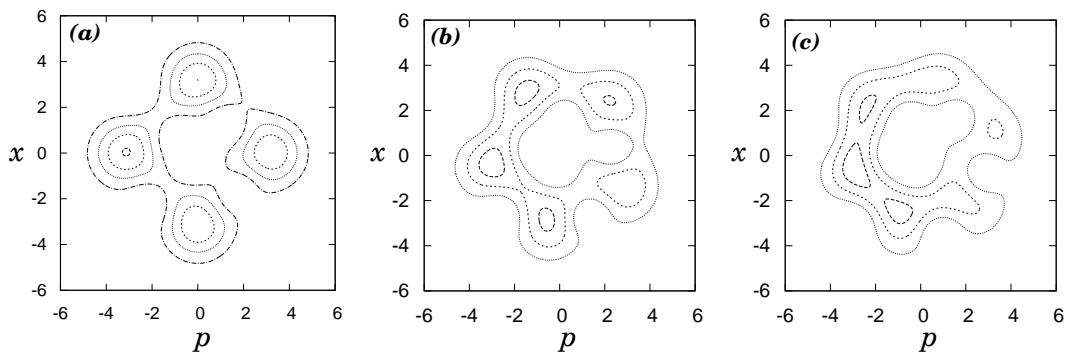}
\caption{Husimi function of the states at (a) $t=T_{\rm rev}/4$, (b) $t=T_{\rm rev}/5$, and (c)$ t=T_{\rm rev}/6$ for an initial coherent state with $\nu=|\alpha|^2=5$.}
\label{Husimim0}
\end{figure*}

At fractional revival times, the time evolved state\\ $\ket{\psi(\pi/q \chi)}$  in Eq. (\ref{superpositionofcs})  is  a superposition of finite number of macroscopically distinguishable coherent states.   A convenient way to visualize these coherent states are by plotting the quasi probability distributions like  Husimi $Q$-function or Wigner function of these states in the phase-plane.  All the coherent states in the  superposition given in  Eq. (\ref{superpositionofcs})  have the same amplitude $|\alpha|$ and each  coherent state falls regularly on a circle of radius $\modu{\alpha}$. The Husimi $Q$-function for a state $\ket{\psi}$
is defined as 
\begin{equation}
Q(x,p)=\frac{1}{\pi}\modu{\int_{-\infty}^{\infty}dx^\prime~\psi_{\beta}(x^\prime)~\psi(x^\prime)}^2,
\end{equation}
where
\begin{equation}
\psi_{\beta}(x^\prime)=\pi^{-1/4}\exp\left[-\frac{(x^\prime-x)^2}{2}+ip(x^\prime-\frac{x}{2})\right]
\end{equation}
and  $\psi (x^\prime)$ are the position representation of the coherent state $\ket{\beta}$ and $\ket{\psi}$ respectively.
 An approximate calculation \cite{miranowicz} based on Husimi $Q$-function shows that the maximum number of well distinguished states that can be obtained for a given field strength $\modu{\alpha}^2$ is 
	\begin{equation}
		N_{max}\cong 2\pi\modu{\alpha}/2\sqrt{\ln 10}.
		 \label{Nmax}
	\end{equation}
 For instance, for  $\modu{\alpha}^2=5$,  $N_{max}\cong 4.62$ and  maximum number of well distinguished states is five (rounding $N_{max}$ to highest integer value). Contour plots of Husimi function   given in Fig. \ref{Husimim0} at fractional revival times $t=T_{\rm rev}/4$, $t=T_{\rm rev}/5$ and $t=T_{\rm rev}/6$ verifies this result. We find that    highest order of fractional  revival that can be observed in the entropy plot is   related  to the value of $N_{max}$. For example, the highest order of fractional revival that can be seen in the entropy plot is five for  $\modu{\alpha}^2=5$ and it is evident  from the dotted curve in Fig. \ref{m0nu510}. 
There are  $9$ well distinguished local minima in this case and they corresponds to $5, 4, 3$ and 2-sub-packet fractional revival times. (It was mentioned earlier, see Eq. (\ref{frtime}),  that $q-$sub-packet fractional revival  occurs at $t/T_{\rm rev}=p/q$.)
When the   field strength $\modu{\alpha}^2$ increases  the radius of the circle in phase-plane increases and  higher orders of fractional revivals are captured in Fig. \ref{m0nu510}. This is  evident in  the dashed and dotted   curves  corresponds to field strengths $\nu=10$ and $20$, respectively  in Fig. \ref{m0nu510}.

\begin{figure}[htpb]
\centering
\includegraphics[scale=0.65]{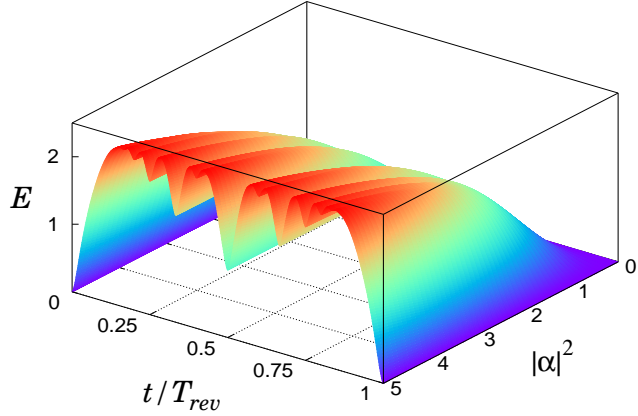}
\caption{Entanglement as a function of time $t/\trev$ and field strength $\modu{\alpha}^2$ for initial coherent state. Entanglement goes to zero at any instant when $\modu{\alpha}^2\rightarrow 0$.  For a given value of $\modu{\alpha}^2$, the entanglement attains a constant highest value   during collapse of wave packets and the highest  value of entanglement increases with increase in the field strength $\modu{\alpha}^2$.}
\label{surf-m0}
\end{figure} 

The entanglement attains a maximum  value  (not to be confused with {\it maximally entangled state} in $q$ dimension) when the wave packet is  completely  collapsed during the evolution. This is due to the fact that the state at the instant  of collapse is having high degree of  nonclassicality.  Complete collapse of the state occurs at a few  instants  of time between the revivals and   the maximum value of entanglement is same at these instants   for a given field strength $|\alpha|^2$.  We denote the maximum value of entanglement by $E_{max}$ and  $E_{max}\cong 2.37$ ebits for $\nu=|\alpha|^2=5$ (see Fig. \ref{m0nu510}).   For $\nu=10$ and $20$, $E_{max}$ is $2.90$ and $3.42$ ebits, respectively.    Fig. \ref{surf-m0} shows the variation of entanglement as a function time $t/\trev$ and field strength $\modu{\alpha}^2$. The maximum value ($E_{max}$) of entanglement  is  constant during the evolution  for a given value of $\nu$ and it increases with increase in $\nu$ value. The value of entanglement becomes zero at any instants of time when $\modu{\alpha}^2\rightarrow 0$ because in this case  both the arms of the beam splitter contains vacuum state $\ket{0}$ which is a classical state. 

\section{Entanglement dynamics of an  initial $m$-photon-added coherent state}
We next consider the entanglement properties of initial photon-added coherent state  (PACS) which is a nonclassical state \cite{agarwal}. Photon-added coherent states $\ket{\alpha,m}$ are excited coherent states:
	\begin{equation}
		\ket{\alpha,m}=N_{\alpha,m}\, {a^\dag}^{m}\ket{\alpha},
		\label{PACS}
	\end{equation}
where $m$ is the number of photons added to the coherent field $\ket{\alpha}$, also called the photon excitation number,   and $N_{\alpha,m}$ is the normalization constant. Setting $m=0$ in Eq. (\ref{PACS}) will retrieve the coherent state $\ket{\alpha}$. It has been shown that degree of  non-classicality of $m$-photon-added coherent states increases with increase in $m$ \cite{sudheesh2} based on a measure defined using the volume of the negative part of the Wigner function \cite{kenfack}.  It is shown that beam splitting of PACS with vacuum produces an entangled state and the amount of entanglement in such state increases with increase in photon excitation number $m$ \cite{berrada}. 

The time evolution of the initial photon-added coherent states through Kerr medium gives
	\begin{eqnarray}
		\ket{\psi(t)}&=&N_{\alpha,m}\,e^{-\modu{\alpha}^2/2}
		 \sum_{n=0}^{\infty}\frac{\alpha^n\sqrt{(n+m)!} }{n!}\nonumber \\
		&&\times e^{-i\chi t (n+m)(n+m-1)}\ket{n+m}.
	\end{eqnarray}
The photon-added coherent states   exhibit  the revival and fractional revivals at the same instants as that of an initial coherent state $\ket{\alpha}$ \cite{sudheesh3}. Beam splitting of the $\ket{\psi(t)}$ with vacuum $\ket{0}$ results in the entangled output state of the form,
	\begin{eqnarray}
		\ket{\Phi(t)}&=N_{\alpha,m}e^{-\modu{\alpha}^2/2}
\sum_{n=0}^\infty \frac{\alpha^n}{n!}2^{-(n+m)/2}\nonumber\\
&\times\exp\left[-i\chi t (n+m)(n+m-1)\right]\nonumber \\
&\times \sum_{p=0}^{n+m}{{n+m}\choose{p}}\sqrt{p!(n+m-p)!}\ket{p}_c\ket{n+m-p}_d.
\label{phi_PACS}
	\end{eqnarray}	
\begin{figure}[h]
\centering
\includegraphics[scale=0.47]{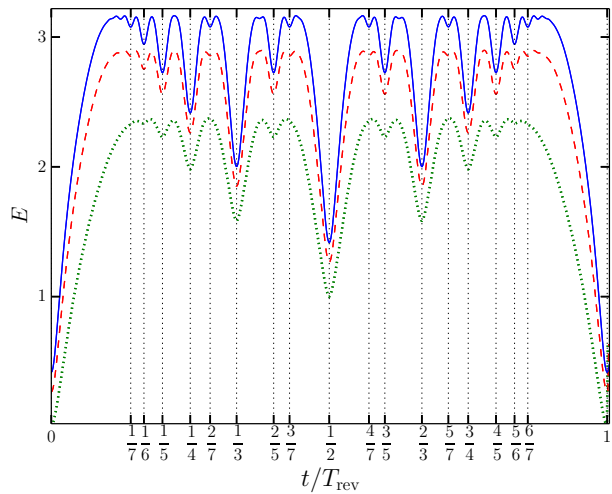}
\caption{Entanglement as a function of time $t/\trev$ for initial $m$-photon-added coherent states for $m=5$ (dashed) and $m=10$ (solid). The dotted curve is for an initial coherent state for reference. Here  $\modu{\alpha}^2=5$ .  Local minima of the curves indicate the instants of  fractional revivals.}
\label{nu5m0510}
\end{figure}
Figure \ref{nu5m0510} shows the plot of entanglement versus time for the initial states $\ket{\alpha,5}$  and $\ket{\alpha,10}$ with  $|\alpha|^2=5$. Entanglement for  an initial coherent state is also plotted in the same figure  for reference. For a given value of $\modu{\alpha}^2$  the entanglement for initial PACS at any  time $t$ is greater than that of an initial coherent state. Entanglement of m-photon-added coherent state is always greater than the entanglement of (m-1)-photon-added coherent state at any instants of time.

\begin{figure*}[htpb]
\centering
\includegraphics[scale=0.6]{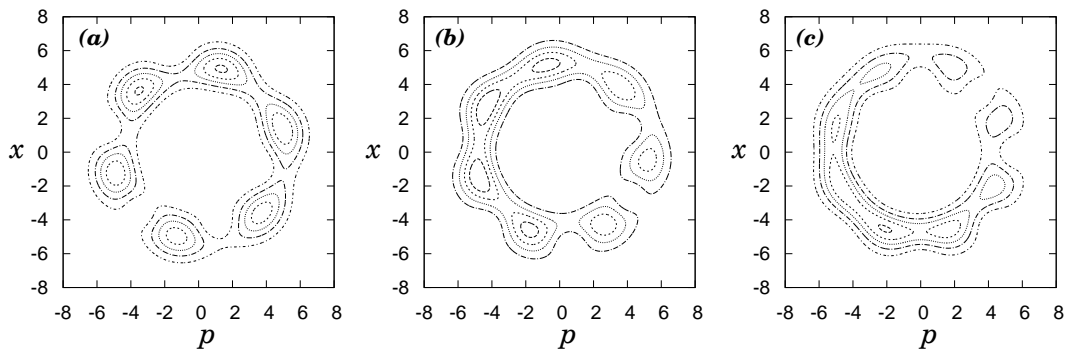}
\caption{Husimi $Q$-function of the states at (a) $t=T_{\rm rev}/6$, (b) $t=T_{\rm rev}/7$, and (c)$ t=T_{\rm rev}/8$ for an initial $5$-photon-added coherent state with $|\alpha|^2=5$.}
\label{Husimim5}
\end{figure*}

 The fractional revivals are indicated by the local minima of the plot. 
For a given value of $|\alpha|^2\,\, (=5$, in the given plot) the highest order of fractional revival seen in the entropy plot increases with increase in $m$. Calculation based on Husimi $Q$-function shows that the maximum number of well distinguished $m$-photon-added coherent states that can be obtained for a given field strength $|\alpha|^2$ depends on the photon excitation number $m$. Our numerical calculation shows that for an initial $5$-photon-added coherent state $N_{\rm max}=6.2$ and for an initial $10$-photon-added coherent state $N_{\rm max}=8.0$ with $|\alpha|^2=5$.  The maximum number of well distinguished states in Husimi $Q-$function plots are  $7$ and
$8$ for  $m=5$ and $10$, respectively (rounding  $N_{\rm max}$ to  the highest integer). 
A contour plot (Fig. \ref{Husimim5})  of the Husimi $Q$-function for a $5$-photon-added coherent state   are at fractional revival times $t=T_{\rm rev}/6$, $t=T_{\rm rev}/7$ and $t=T_{\rm rev}/8$.  Again, we find that the   highest order of fractional  revival that can be observed in the entropy plot is  related  to the value of $N_{max}$. The highest order of fractional revival that can be seen in the entropy plot (Fig. \ref{nu5m0510}) are $7$ and $8$ for the  initial $5$ and $10$-photon-added coherent states, respectively. 
Entanglement attains  a maximum value $E_{max}$ when the wave packet is completely collapsed during the evolution as in the case of the initial coherent state. The maximum value of entanglement  increases with increase in photon excitation number $m$ and the value of  $|\alpha|^2$. The entanglement at any time $t$ is much greater than the entanglement obtained by the direct beam splitting of the initial photon-added coherent states with the vacuum. Thus the time evolution of the initial wave packet in the medium generates a very large amount of entanglement.

\begin{figure}
\centering
\includegraphics[scale=0.65]{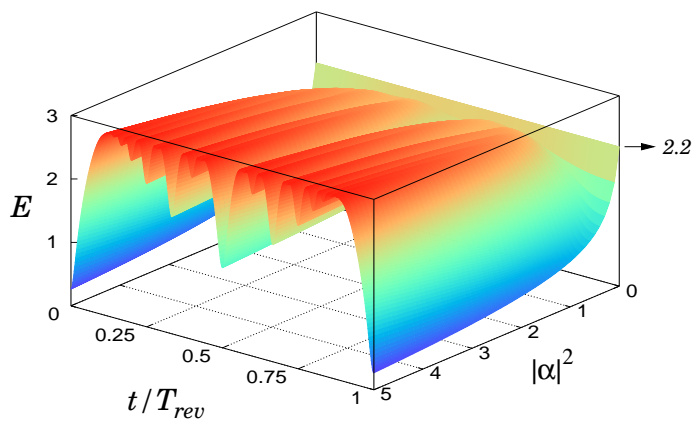}
\caption{Entanglement as a function of time $t/\trev$ and $\modu{\alpha}^2$ for an initial $5$-photon added coherent state.  Maximum value of entanglement  increases with increase in the field strength $\modu{\alpha}^2$.}
\label{surf-m5}
\end{figure}

Fig. \ref{surf-m5} gives the variation of entanglement as a function of time $t/\trev$ and field strength $\modu{\alpha}^2$ for initial $5$-photon added coherent state. The value of entanglement becomes $2.2$ ebits at any instants of time when $\modu{\alpha}^2\rightarrow0$.  This value corresponds to the entanglement of state obtained using the beam splitting of    Fock state $\ket{5}$ with vacuum. The reason is that a  $m$-photon-added coherent state reduces to the Fock state $\ket{m}$ when $\modu{\alpha}^2\rightarrow0$. 

\section{Effect of decoherence on the entangled states}
In this section,  we study the effect of decoherence on the entanglement between the output modes of the beam splitter due to photon absorption by an  environment. We assume that both of the output modes of the beam splitter interact independently with an external environment which is modelled  as a collection of infinite number of harmonic oscillators. In the  rotating wave approximation, the interaction of the  output modes $c$ and $d$ of the beam splitter with the environment modes $e_j$  can be described by the Hamiltonian $H_{int}=\sum_{j}\gamma_k\left(a_k^\dag e_j+e_j^\dag a_k\right)$, where  $\gamma_k$ is the coupling strength of the mode $a_k$ (mode $c$ or $d$)  with the environment.  Such decoherence process can be effectively described by the quantum mechanical master equation. Within Born-Markov approximation, the density matrix $\rho_{cd}$ of bipartite field modes at the output of the beam splitter obeys zero temperature master equation:
	\begin{equation}
		\frac{d \rho_{cd}}{d \tau}=-i\sum_{k=1}^{2} \omega_k \left[a_k^\dag a_k,\rho_{cd}\right]+\sum_{k=1}^{2}\gamma_k \left(2 a_k \rho_{cd} a_k^\dag-a_k^\dag a_k \rho_{cd} -\rho_{cd} a_k^\dag a_k  \right). \label{master}
	\end{equation}
	The first term in the right hand side of Eq. (\ref{master}) is the unitary evolution  and the second term is  the dissipative evolution. Following the procedure for solving  master equation of coupled nonlinear oscillator using  thermofield dynamics notation\cite{chaturvedi}, we find  	the matrix elements of $\rho_{cd}(\tau)$ in the Fock basis, 
\begin{equation}
\bra{m_1,m_2}\rho_{cd}(\tau)\ket{n_1,n_2}=\sum_{p_1,p_2=0}^{\infty} R_1 R_2 \bra{m_1+p_1,m_2+p_2}\rho_{cd}(\tau=0)\ket{n_1+p_1,n_2+p_2},\label{master-solution}
\end{equation}
where
\begin{equation}
R_j={{{m_j+p_j}\choose{p_j}}}^{1/2} {{{n_j+p_j}\choose{p_j}}}^{1/2} {\left(1-\exp({-2\gamma_j \tau})\right)}^{p_j} \exp({-\gamma_j \tau(m_j+n_j)})
\end{equation}
for $j=1, 2$. The density matrix 
\begin{equation}
\rho_{cd}(\tau=0)=\ket{\Phi(t)}\bra{\Phi(t)},
\end{equation}	
where $\ket{\Phi(t)}$ is the entangled output state given in Eq.  (\ref{phi_PACS}) for an initial $m$-photon-added coherent state. 

The state  $\rho_{cd}(\tau)$ is a mixed state for $\tau >0$ and we choose   logarithmic negativity\cite{vidal} to quantify the entanglement. The logarithmic negativity is defined as
\begin{equation}
E_N=\log_2 \parallel \rho_{cd}^{T_k} \parallel,
\end{equation}
where $\parallel \cdot \parallel$ denotes the trace norm operation, which  is equal to the sum of the absolute values of eigen values  for a Hermitian operator, and $\rho^{T_k}$ with $k=1$ (or $2)$ represents the partial transpose of $\rho_{cd}$ with respect to mode $c$  (or $d)$. 

\begin{figure}
\centering
\includegraphics[scale=0.55]{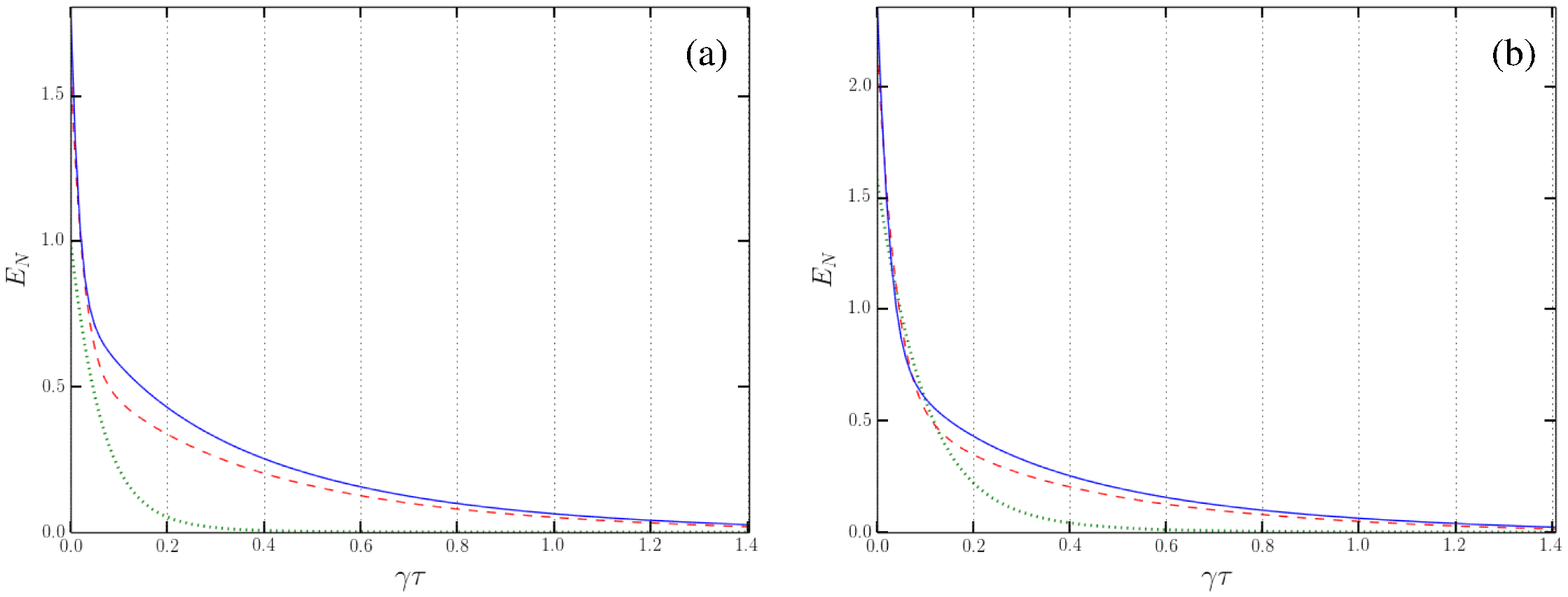}
\caption{Decay of entanglement as a function of $\gamma \tau$ for  of the state $\ket{\Phi(t)}$ at  (a) $t=\trev/2$ and  (b) $t=\trev/3$. The solid and dashed curves corresponds to the  photon-added coherent states  $\ket{\alpha, 10}$ and $\ket{\alpha, 5}$, 
respectively. Here, $\modu{\alpha}^2=5$.}
\label{Envstau1}
\end{figure}

\begin{figure}
\centering
\includegraphics[scale=0.55]{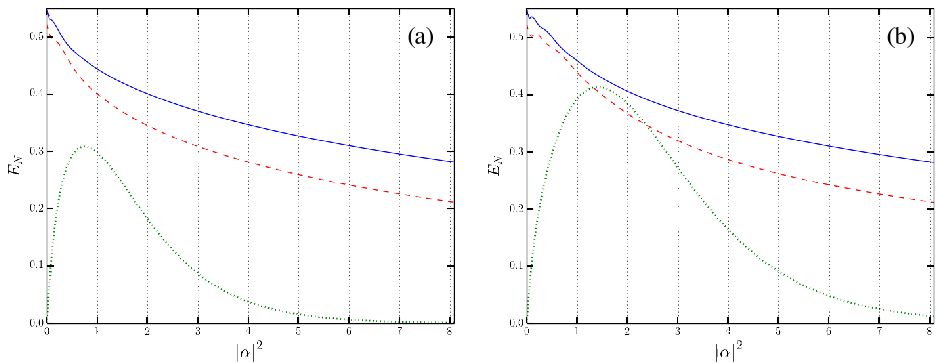}
\caption{Decay of entanglement as a function of $|\alpha|^2$ for  of the state $\ket{\Phi(t)}$ at  (a) $t=\trev/2$ and  (b) $t=\trev/3$. The solid and dashed curves corresponds to the  photon-added coherent states  $\ket{\alpha, 10}$ and $\ket{\alpha, 5}$, 
respectively. Here, $\gamma\tau=0.3$.}
\label{Envsalpha}
\end{figure}

Here, we present our results obtained for the decoherence of the bipartite states  $\ket{\Phi(t)}$ for  the initial $m$-photon-added coherent states
at the instants of fractional revivals.  Decay of entanglement due to photon absorption losses
 for an initial coherent state (m=0 case in  Eq.  (\ref{phi_PACS})) is  already discussed in\cite{vanEnk2}. Time evolved matrix elements of the states $\rho_{cd}(\tau)$ are calculated using   Eq. (\ref{master-solution}) and logarithmic negativity $E_N$ 
is calculated numerically. Without loss of generality, we assume that the coupling constants $\gamma_1=\gamma_2=\gamma=0.1$.  In Fig. \ref{Envstau1}, we plotted the  entanglement $E_N$  of the state $\rho_{cd}(\tau)$  as a function of  $\gamma\tau$.
The Figures \ref{Envstau1} (a) and  (b)  correspond to the  initial state $\rho_{cd}(\tau=0)$ at  fraction revival instants $t=\trev/2$ and  $\trev/3$, respectively. In each figure, the solid and dashed curves correspond to the initial photon-added coherent  states  $\ket{\alpha, 10}$ and $\ket{\alpha, 5}$, 
respectively.  We also plotted (dotted lines) the entanglement versus $\gamma \tau$ for an  initial coherent state   for reference.  It is evident from the Fig. \ref{Envstau1} that   the entanglement  decreases with time and becomes zero for large $\gamma\tau$ irrespective of the initial states considered. The figures  shows that entanglement decay of the initial $m$-photon-added coherent states are slower 
when it is  compared with the initial coherent state.   
We have also plotted, see Fig. \ref{Envsalpha},   entanglement $E_N$ versus $|\alpha|^2$ at $\gamma \tau=0.3$ for the cases discussed above. We find that entanglement of initial photon-added coherent states are decaying very slowly  as a function of $|\alpha|^2$ when it is compared with initial coherent state. Thus, the states $\ket{\Phi(t)}$ generated at the output of the beam splitter using initial $m$-photon-added coherent states (nonclassical states) are  more robust against decoherence than those formed by an initial coherent state (classical state).

\section{Conclusions}

Entanglement dynamics of initial classical and nonclassical states are studied and compared  using a beam splitter with a Kerr medium in one of its  input mode.  Arbitrary large amount of entanglement are generated at the instants of collapses of wave packets during the evolution in the medium. Dynamics of entanglement shows local minima at the instants of fractional revivals and entanglement increases with the increase in the field strength.  The highest order of fractional revivals that can be observed in the entanglement plots are explained using the Husimi $Q$-function. The value of entanglement increases with the increase  in the degree of nonclassicality of the initial $m$-photon-added coherent state.  The highest order of fractional revivals than can be seen in the entropy plot depends on both  the photon excitation number of the initial $m$-photon-added coherent states and  the value of  $|\alpha|^2$. 
We  showed   that the states  generated at the output of the beam splitter using initial $m$-photon-added coherent   states are  more robust against decoherence, due to photon absorption
by an environment, than those formed by an initial coherent state.
In view of the recent experimental demonstration of quantum state collapse and revival of an initial coherent state in a Kerr medium \cite{gerhard}  and the production of single-photon-added coherent states \cite{zavatta}, our theoretical results in this paper may be experimentally tested and verified.

\end{document}